\documentclass[corr]{new_tlp_corr}
\usepackage{amsmath}
\usepackage{amsfonts}
\usepackage{amssymb}
\usepackage{listings}
\usepackage{galois}
\usepackage{url}
\usepackage{tikz}

\newcommand{\headcost}{\varphi} % was \beta 
\newcommand{\litcost}{\omega}   % was \delta 
\newcommand{\constantone}{c_{1}} % was \alpha
\newcommand{\constanttwo}{c_{2}} % was \beta 

\newcommand{\basetype}{\eta} % was \alpha

\newcommand{\ciaopp}{CiaoPP}
\newcommand{\ciao}{Ciao}

\def\imp{\hbox{${\tt \ :\!-\ }$}}

\title[Resource Usage Analysis via Abstract Interpretation Using Sized Types ]
      {
      Resource Usage Analysis of Logic Programs\\ 
      via Abstract Interpretation Using Sized Types 
\thanks{This research was supported in part by projects EU 
  FP7 318337 \emph{ENTRA}, Spanish MINECO 
  TIN2012-39391 \emph{StrongSoft} and TIN2008-05624
  \emph{DOVES}, and Madrid TIC/1465 \emph{PROMETIDOS-CM}.} 
      }

\author[A. Serrano et al.]{
       A. SERRANO$^{1}$\thanks{A. Serrano performed this work during his former affiliation to the IMDEA Software Institute.}
       ~~ P. LOPEZ-GARCIA$^{2,3}$
       ~~ M. V. HERMENEGILDO$^{3,4}$
 \\
 \\
   $^1$Dept. of Information and Computing Sciences, Utrecht University \\
   \email{A.SerranoMena@uu.nl} \\
   $^2$IMDEA Software Institute \\
   \email{pedro.lopez@imdea.org, manuel.hermenegildo@imdea.org} \\
   $^3$Spanish Council for Scientific Research  (CSIC) \\
   $^4$Technical University of Madrid (UPM) \\
   \email{herme@fi.upm.es}\\ 
   \vspace*{-1mm}
}
       
\jdate{January 2014}
\pubyear{2014}
\pagerange{\pageref{firstpage}--\pageref{lastpage}}

\begin{document}

\label{firstpage}

\maketitle

\begin{abstract}

We present a novel general resource analysis for logic programs based
on sized types.  Sized types are representations that incorporate
structural (shape) information and allow expressing both lower and
upper bounds on the size of a set of terms and their subterms at any
position and depth.  They also allow relating the sizes of terms and
subterms occurring at different argument positions in logic
predicates.  Using these sized types, the resource analysis can infer
both lower and upper bounds on the resources used by all the
procedures in a program as functions on input term (and subterm)
sizes, overcoming limitations of existing resource analyses and enhancing their
precision. Our new resource analysis has been developed within the
abstract interpretation framework, as an extension of the sized types
abstract domain, and has been integrated into the \ciao\ preprocessor,
\ciaopp.  The abstract domain operations are integrated with the
setting up and solving of recurrence equations for inferring both size
and resource usage functions.  We show that the analysis is an
improvement over the previous resource analysis present in
\ciaopp\ and compares well in power to state of the art systems.
% for
%functional programs.
\end{abstract}

\section{Introduction}

\emph{Resource usage analysis} infers the aggregation of some
numerical properties (named \emph{resources}), like memory usage, time
spent in computation, or bytes sent over a wire, throughout the
execution of a piece of code.
The expressions giving the usage of resources are usually functions 
of the sizes of some input arguments to procedures.

Our starting point is the methodology outlined
% For CoRR
% by~\cite{granularity,caslog,low-bounds-ilps97},
by \\ \cite{granularity,caslog,low-bounds-ilps97},
characterized by the setting up of recurrence equations. In that
methodology, the size analysis is the first of several other analysis
steps that include, e.g., cardinality analysis (that infers lower and upper
bounds on the number of solutions computed by a predicate), and which
ultimately obtain the resource usage bounds. One drawback of these
proposals, as well as most of their subsequent derivatives, is that
they are able to cope with size information about subterms in a
very limited way. This is an important limitation, which causes the 
analysis to infer trivial bounds for a large class of programs.  For
example, consider a predicate which computes the factorials of a list:

\vspace{-2mm}
\noindent
\begin{small}
\begin{tabular}{p{6.3cm}p{6cm}}
\begin{lstlisting}
% listfact(+L, -FL).
listfact([],   []).
listfact([E|R],[F|FR]) :-
	fact(E, F),
	listfact(R, FR).
	
\end{lstlisting} & 
\begin{lstlisting}
% fact(+N, -F).
fact(0,1).
fact(N,M) :- N1 is N - 1,
             fact(N1, M1),
             M is N * M1.
\end{lstlisting}
\end{tabular}
\end{small}

\vspace{-5mm}

\noindent Intuitively, the best bound for the running time of this
program for a list
$L$ is $\constantone + \sum_{e \in L} \left( \constanttwo + time_{fact}(e) \right)$,
where $\constantone$ and $\constanttwo$ are constants related to unification
and calling costs. But with no further information, the upper
bound for the elements of $L$ must be $\infty$ to be on the safe side,
and then the returned overall time bound must also be $\infty$.
In a previous paper~\cite{sized-types-iclp2013} we focused on a
proposal to improve the size analysis based on \emph{sized
  types}. 
While in that paper we already hinted at the fact that the application
of our sized types in resource analysis could result in considerable
improvement,
% allowing the inference of equal or better resource usage
% bounds than comparable state of the art systems, but 
no description was provided of the actual resource analysis. 
This paper is complementary 
% to the mentioned previous paper~\cite{sized-types-iclp2013},
% focusing instead on the \emph{application} of sized types to build a
% powerful resource analysis. In particular,
and fills this gap by describing a new resource usage analysis 
% with two novel aspects. Firstly, it 
that can \emph{take advantage of the new
  information contained in sized types}. 
Furthermore, the resource analysis we propose is \emph{based fully on
  abstract interpretation}.  Previously, the auxiliary
analyses % were developed using
used this technique,
% abstract interpretation, 
% whereas 
but the core resource analysis
% was outside this framework. 
did not use it directly. Our approach formulates the 
resource analysis as an \emph{abstract domain} that can be integrated 
within a standard, parametric abstract interpreter.
% and inherit its capabilities.
In particular, % we integrate the analysis as a domain 
we integrate it into the 
PLAI abstract interpretation framework~\cite{ai-jlp,inc-fixp-sas}
% This allows us to integrate the analysis 
of % the 
\ciaopp{},
%  system, which gives us 
obtaining features such as
\emph{multivariance}, efficient fixpoints, and assertion-based
verification and user interaction for free. We also perform an 
% performance 
assessment of the accuracy and efficiency
of the resulting overall system.

In Section \ref{sec:overview} we give a high-level view of the
approach.
%In the next two sections we review the abstract
%interpretation framework and the sized types approach to size
%analysis.
In the following section we review the abstract interpretation
approach to size analysis using sized types.
Section~\ref{sec:resources} gets deeper into the resource usage
analysis, our main contribution. Experimental results are shown in
Section~\ref{sec:results}. Finally we review some related work and
discuss future directions.

\section{Overview of the Approach}
\label{sec:overview}

We give now an overview of our approach to resource usage analysis,
and present the main ideas in our proposal using the classical
\texttt{append/3} predicate as a running example:

\begin{small}
\begin{lstlisting}
append([],    S, S).
append([E|R], S, [E|T]) :- append(R, S, T).
\end{lstlisting}
\end{small}

\noindent
The process starts by performing the regular type analysis present in
the \ciaopp{} system~\cite{eterms-sas02}. In our example, the system
infers that for any call to the predicate \texttt{append(X, Y, Z)}
with \texttt{X} and \texttt{Y} bound to lists of numbers and
\texttt{Z} a free variable, if the call succeeds, then \texttt{Z} also
gets bound to a list of numbers. The set of ``list of numbers'' is
represented by the regular type $listnum$, defined as follows:
\begin{lstlisting}
listnum := [] | [num | listnum].
\end{lstlisting}
% We should use Ciao syntax for programs and -> type rules for math,
% right? 
% listnum -> [] | .(num, listnum)

From this regular type definition, sized type schemas are derived. 
The sized type schema $listnum\text{-}s$ is derived from
$listnum$. This schema corresponds to a list whose length is
between $\alpha$ and $\beta$, containing numbers between
$\gamma$ and $\delta$.
$$listnum\text{-}s \to listnum^{(\alpha,\beta)}(num^{(\gamma, \delta)})$$

From now on, in the examples we will use $ln$ and $n$ instead of
$listnum$ and $num$ for the sake of conciseness. The next phase
involves relating the sized types of the different arguments to the
\texttt{append/3} predicate using recurrence (in)equations.  Let $size_X$
denote the sized type schema for argument \texttt{X} in a
call \texttt{append(X, Y, Z)} (from the regular type inferred
by a previous analysis). We have that $size_X$ denotes $ln^{(\alpha_X,
  \beta_X)}(n^{(\gamma_X, \delta_X)})$.
% be the sized type of a list $X$ of numbers.  
Similarly, the sized type schema for the output argument \texttt{Z} is
$ln^{(\alpha_Z, \beta_Z)}(n^{(\gamma_Z,
  \delta_Z)})$, denoted by $size_Z$. We are interested in
expressing bounds on the length of the output list \texttt{Z} and the
values of its elements as a function of size bounds for the input lists
\texttt{X} and \texttt{Y} (and their elements). For this, we
set up a system of inequations.  For instance, the inequations that are
set up to express a lower bound on the length of the output argument
\texttt{Z}, denoted $\alpha_Z$, as a function on
the size bounds of the input arguments \texttt{X} and \texttt{Y}, and
their subarguments ($\alpha_X,\ \beta_X, \ \gamma_X, \ \delta_X,
\ \alpha_Y, \ \beta_Y, \ \gamma_Y$, and $\delta_Y$) are:

\vspace{-2mm}
\begin{small}
$$
\alpha_Z
\begin{pmatrix}
\alpha_X,\beta_X,\gamma_X,\delta_X, \\
\alpha_Y,\beta_Y,\gamma_Y,\delta_Y
\end{pmatrix}
\geq   % -> for lower bounds.
% \leq % -> for upper bounds.
\begin{cases}
\alpha_Y & 
\text{if } \alpha_X = 0
\\
1 + \alpha_Z
\begin{pmatrix}
\alpha_X-1,\beta_X-1, \gamma_X,\delta_X, \\
\alpha_Y,\beta_Y,\gamma_Y,\delta_Y
\end{pmatrix} & 
\text{if } \alpha_X > 0
\end{cases}$$
\end{small}

% for the upper bound on the elements, would read:

\noindent
Note that in the recurrence inequation set up for the second clause of
\texttt{append/3}, the expression $\alpha_X-1$ (respectively
$\beta_X-1$) represents the size relationship that a lower
(respectively upper) bound on the length of the list in the first
argument of the recursive call to \texttt{append/3} is one unit less
than the length of the first argument in the clause head.

As the number of size variables grows, the set of inequations becomes too
large. Thus, we propose a compact representation,
which allows us to grasp all the relations in one view.
The first change in
our proposal is to write the parameters to size functions directly as
sized types. Now, the parameters to the $\alpha_Z$ function are the
sized type schemas corresponding to the arguments \texttt{X} and
\texttt{Y} of the \texttt{append/3} predicate:

\vspace{-2mm}
\begin{small}
$$
\alpha_Z
\begin{pmatrix}
ln^{(\alpha_X, \beta_X)}(n^{(\gamma_X, \delta_X)}) \\
ln^{(\alpha_Y, \beta_Y)}(n^{(\gamma_Y, \delta_Y)})
\end{pmatrix}
\geq   % -> for lower bounds.
% \leq % -> for upper bounds.
\begin{cases}
\alpha_Y &
\text{if } \alpha_X = 0
\\
1 + \alpha_Z
\begin{pmatrix}
ln^{(\alpha_X - 1, \beta_X - 1)}(n^{(\gamma_X, \delta_X)}) \\
ln^{(\alpha_Y, \beta_Y)}(n^{(\gamma_Y, \delta_Y)})
\end{pmatrix} & 
\text{if } \alpha_X > 0
\end{cases}$$
\end{small}

In a second step, we group together all the inequalities of a single
sized type. As we always alternate lower and upper bounds, it is
always possible to distinguish the type of each inequality. We do not
write equalities, so that we do not use the symbol $=$. However, we
always write inequalities of both signs ($\geq$ and $\leq$) for each
size function, since we compute both lower and upper size bounds.  
% to separate the left and hand sides of the inequalities,
Troughout this paper we use a representation using $\lessgtr$ for the symbols
$\geq$ and $\leq$ that are always paired. For example, the
expression
$ln^{(\alpha_X, \beta_X)}(n^{(\gamma_X, \delta_X)})
  \lessgtr
  ln^{(e_1, e_2)}(n^{(e_3, e_4)})$
represents the conjunction of the following size constraints:% \\
\ $\alpha_X \geq e_1, \ \beta_X \leq e_2, \ \gamma_X \geq e_3, \ \delta_X \leq e_4$.
In the implementation, constraints for each variable are kept apart
and solved separatedly.

After setting up the corresponding system of inequations for the
output argument \texttt{Z} of \texttt{append/3}, and solving it, we
obtain the following expression:

\vspace{-2mm}
\begin{small}
$$size_Z\left(
%  ln^{(\alpha_X, \beta_X)}(n_{\langle ., 1 \rangle}^{(\gamma_X, \delta_X)}),
%  ln^{(\alpha_Y, \beta_Y)}(n_{\langle ., 1 \rangle}^{(\gamma_Y, \delta_Y)})
size_X, size_Y
\right)
\lessgtr
  ln^{( \alpha_X + \alpha_Y, \beta_X + \beta_Y )}
   (n^{(\min(\gamma_X, \gamma_Y), \max(\delta_X, \delta_Y))}) $$
\end{small}
\vspace{-4mm}

\noindent that represents, among others, the relation $\alpha_z \geq \alpha_X +
\alpha_Y$ (resp.  $\beta_z \leq \beta_X + \beta_Y$), expressing that a
lower (resp. upper) bound on the length of the output list \texttt{Z},
denoted $\alpha_z$ (resp. $\beta_z$), is the addition of the lower
(resp. upper) bounds on the lengths of \texttt{X} and \texttt{Y}. It
also represents the relation $\gamma_Z \geq \min(\gamma_X, \gamma_Y)$
(resp. $\delta_Z \leq \max(\delta_X, \delta_Y)$), which expresses that
a lower (resp. upper) bound on the size of the elements of the list
\texttt{Z}, denoted $\gamma_z$ (resp. $\delta_z$), is the minimum
(resp. maximum) of the lower (resp. upper) bounds on the sizes of the
elements of the input lists \texttt{X} and \texttt{Y}.

Resource analysis builds upon the sized type analysis and adds
recurrence equations for each resource we want to analyze. Apart from
that, when considering logic programs, we have to take into account
that they can fail or have multiple solutions when executed, so we
need an auxiliary \emph{cardinality analysis} to get correct results.

Let us focus on cardinality analysis. Let $s_L$ and $s_U$ denote
lower and upper bounds on the number of solutions for
\texttt{append/3}. Following the program
structure we can infer:

\vspace{-2mm}
\begin{small}
$$
\begin{array}{rcl}
s_L\left(
  ln^{(0, 0)}(n^{(\gamma_X, \delta_X)}),
  size_Y % ln^{(\alpha_Y, \beta_Y)}(n^{(\gamma_Y, \delta_Y)})
\right) 
& \geq &
  1 \\
s_L\left(
  ln^{(\alpha_X, \beta_X)}(n^{(\gamma_X, \delta_X)}),
  size_Y % ln^{(\alpha_Y, \beta_Y)}(n^{(\gamma_Y, \delta_Y)})
\right) 
& \geq &
  s_L\left(
  ln^{(\alpha_X - 1, \beta_X - 1)}(n^{(\gamma_X, \delta_X)}),
  size_Y % ln^{(\alpha_Y, \beta_Y)}(n^{(\gamma_Y, \delta_Y)})
\right) 
\end{array}$$
\vspace{-3mm}
$$
\begin{array}{rcl}
s_U\left(
  ln^{(0, 0)}(n^{(\gamma_X, \delta_X)}),
  size_Y %  ln^{(\alpha_Y, \beta_Y)}(n^{(\gamma_Y, \delta_Y)})
\right) 
& \leq &
  1 \\
s_U\left(
  ln^{(\alpha_X, \beta_X)}(n^{(\gamma_X, \delta_X)}),
  size_Y %  ln^{(\alpha_Y, \beta_Y)}(n^{(\gamma_Y, \delta_Y)})
\right) 
& \leq &
  s_U\left(
  ln^{(\alpha_X - 1, \beta_X - 1)}(n^{(\gamma_X, \delta_X)}),
  size_Y % ln^{(\alpha_Y, \beta_Y)}(n^{(\gamma_Y, \delta_Y)})
\right) 
\end{array}$$
\end{small}
\vspace{-1mm}

\noindent Since $s_L \leq s_U$, the solution to these inequations must be $(s_L, s_U) = (1,1)$. Thus, we have
inferred that \texttt{append/3} has at least (and at most) one solution:
it behaves like a function.
% is deterministic.
When setting up the equations, we use the result of the non-failure analysis to see that
\texttt{append/3} cannot fail when given lists as arguments. If not,
the lower bound is 0.

Now we move forward to analyzing the number of resolution steps
performed 
by a call to \texttt{append/3} (we will only focus on upper bounds,
$r_U$, for brevity). For the first clause, we know that only one
resolution step is needed, so:

\vspace{-2mm}
\begin{small}
$$r_U\left(
  ln^{(0, 0)}(n^{(\gamma_X, \delta_X)}),
  ln^{(\alpha_Y, \beta_Y)}(n^{(\gamma_Y,
    \delta_Y)}) \right) \leq 1 $$
\end{small}
\vspace{-2mm}
%\vspace{-5mm}

\noindent The second clause performs one
resolution step plus all the resolution steps performed by all possible
backtrackings over the call in the body of the clause. This number
can be bounded as a function of the number of solutions.
Thus, the equation reads:

%\begin{small}
% $$\begin{array}{l}
% r_U\left(
%     ln^{(\alpha_X, \beta_X)}(n_{\langle ., 1 \rangle}^{(\gamma_X, \delta_X)}),
%     ln^{(\alpha_Y, \beta_Y)}(n_{\langle ., 1 \rangle}^{(\gamma_Y, \delta_Y)}) \right) \\
% \qquad \leq 1 + s_U\left(
%   ln^{(\alpha_X - 1, \beta_X - 1)}(n_{\langle ., 1 \rangle}^{(\gamma_X, \delta_X)}),
%   ln^{(\alpha_Y, \beta_Y)}(n_{\langle ., 1 \rangle}^{(\gamma_Y, \delta_Y)}) \right) \\
% \qquad \qquad \times r_U\left(
%   ln^{(\alpha_X - 1, \beta_X - 1)}(n_{\langle ., 1 \rangle}^{(\gamma_X, \delta_X)}),
%   ln^{(\alpha_Y, \beta_Y)}(n_{\langle ., 1 \rangle}^{(\gamma_Y, \delta_Y)}) \right) \\
% \qquad = r_U\left(
%   ln^{(\alpha_X - 1, \beta_X - 1)}(n_{\langle ., 1 \rangle}^{(\gamma_X, \delta_X)}),
%   ln^{(\alpha_Y, \beta_Y)}(n_{\langle ., 1 \rangle}^{(\gamma_Y, \delta_Y)}) \right)
% \end{array}$$
\vspace{-2mm}
\begin{small}
$$\begin{array}{rclcl}
  r_U\left(
    ln^{(\alpha_X, \beta_X)}(n^{(\gamma_X, \delta_X)}),
    size_Y % ln^{(\alpha_Y, \beta_Y)}(n^{(\gamma_Y, \delta_Y)})
  \right) 
  & \leq & 1 & + & 
  s_U\left(
  ln^{(\alpha_X - 1, \beta_X - 1)}(n^{(\gamma_X, \delta_X)}),
  size_Y % ln^{(\alpha_Y, \beta_Y)}(n^{(\gamma_Y, \delta_Y)})
  \right)
\\ & & & \times &
  r_U\left(
  ln^{(\alpha_X - 1, \beta_X - 1)}(n^{(\gamma_X, \delta_X)}),
  size_Y % ln^{(\alpha_Y, \beta_Y)}(n^{(\gamma_Y, \delta_Y)})
  \right)
\\ & = & 1 & + & r_U\left(
  ln^{(\alpha_X - 1, \beta_X - 1)}(n^{(\gamma_X, \delta_X)}),
  size_Y % ln^{(\alpha_Y, \beta_Y)}(n^{(\gamma_Y, \delta_Y)})
  \right)
\end{array}$$
\end{small}
\vspace{-1mm}

\noindent Solving these equations we infer that an upper bound on the number of
resolution steps is the (upper bound on) the length of the input list
\texttt{X} plus one. This is expressed as:

\vspace{-2mm}
\begin{small}
$$r_U\left(
    ln^{(\alpha_X, \beta_X)}(n^{(\gamma_X, \delta_X)}),
    ln^{(\alpha_Y, \beta_Y)}(n^{(\gamma_Y, \delta_Y)}) \right)
  \leq \beta_X + 1$$
\end{small}

\vspace{-8mm} % margins not set correctly
\section{Sized Types Review}
\label{sec:sizedtypes}

As shown in the \texttt{append} example, the 
% (bound) 
variables that we relate in our
inequations come from sized types, which are ultimately derived from
the regular types previously inferred for the program. Among several representations of
regular types used in the literature, we use one based on
\emph{regular term grammars}, equivalent to~\cite{Dart-Zobel} but with
some adaptations.  A \emph{type term} is either a \emph{base type}
$\basetype_i$ (taken from a finite set), a \emph{type symbol} $\tau_i$
(taken from an infinite set), or a term of the form $f(\phi_1, \dots,
\phi_n)$, where $f$ is a $n$-ary function symbol (taken from an
infinite set) and $\phi_1, \dots, \phi_n$ are \emph{type terms}.  A
\emph{type rule} has the form $\tau \to \phi$, where $\tau$ is a
\emph{type symbol} and $\phi$ a \emph{type term}. A \emph{regular term
  grammar} $\Upsilon$ is a set of \emph{type rules}.

% To devise the abstract domain we focus specifically on the PLAI
% algorithm of \cite{ai-jlp}, which uses the
% % Repetition -PLG
% % generic 
% \textsc{and-or} tree abstraction of
% \cite{DBLP:journals/jlp/Bruynooghe91}, 
% but implements it implicitly and efficiently using a memo table-based
% fixpoint. The PLAI procedure is \emph{generic} and goal
To devise the abstract domain we focus specifically on the
% Repetition -PLG
% generic 
% For CoRR
% PLAI~\cite{abs-int-naclp89,ai-jlp} framework, 
PLAI \\ \cite{abs-int-naclp89,ai-jlp} framework, 
% which is also 
% the basis of the abstract analyzer present 
integrated within 
% For CoRR
% \ciaopp~\cite{hermenegildo11:ciao-design-tplp} (see~\ref{sec:absframe}),
\ciaopp \cite{hermenegildo11:ciao-design-tplp} (see~\ref{sec:absframe}),
where we have incorporated our implementation.
% of the proposed analysis.
The PLAI algorithm abstracts execution \textsc{and-or} trees similarly
to~\cite{DBLP:journals/jlp/Bruynooghe91} but represents the abstract
executions \emph{implicitly} and computes fixpoints efficiently using
memo tables, dependency tracking, etc.
% of the \textsc{and-or} trees 
% procedure
% It is \emph{generic} and goal dependent: 
It takes as input a pair $(L,\lambda_c)$ representing an entry point
(predicate) along with an abstraction of the call patterns (in the
chosen \emph{abstract domain}) and produces an
abstraction % $\lambda_o$
which overapproximates information at all program points (for all
procedure versions).

The formal concept of \emph{sized type} is an abstraction of a set of
Herbrand terms which are a subset of some regular type $\tau$ and meet
some lower- and upper-bound size constraints on the number of
\emph{type rule applications} needed to generate the terms. A grammar for the new sized types
follows:

\vspace{-2mm}

\medskip
\noindent
\figrule
\begin{tabular}{rclr}
%\multicolumn{4}{c}{\emph{Sized type syntax}} \\
%\hline
\emph{sized-type} & $::=$  & $\basetype^{bounds}$ & $\basetype$ base type \\
                  & $|$    & $\tau^{bounds}(\textit{sized-args})$ & $\tau$ recursive type symbol \\
                  & $|$    & $\tau(\textit{sized-args})$ & $\tau$ non-recursive type symbol \\
\emph{bounds}     & $::=$  & $nob$ \ $|$ \ $(n, m)$  & $n,m \in \mathbb{N}, m \geq n$ \\
\emph{sized-args} & $::=$  & $\epsilon$  \ $|$ \ \emph{sized-arg}, \ \emph{sized-args}   \\
\emph{sized-arg}  & $::=$  & $\textit{sized-type}_{position}$ \\
\emph{position}   & $::=$  & $\epsilon$ \ $|$ \ $\langle f, n \rangle$ & $f$ functor, $0 \leq n \leq$ arity of $f$ \\
%\hline
\end{tabular}
\noindent\figrule

\vspace{-4mm}

\medskip
\noindent However, in our abstract domain we need to refer to sets of
sized types which satisfy certain constraints on their bounds. For
that purpose, we introduce \emph{sized type schemas}: a schema is just
a sized type with variables in bound positions, i.e., where $n$ and
$m$ in the pair $(n, m)$ defining the symbol \emph{bounds} in the
grammar above are variables (called bound variables), along with a set of constraints over
those variables. We call such variables \emph{bound variables}. We
will denote $sized(\tau)$ the sized type schema corresponding to a
regular type $\tau$ where all the bound variables are fresh.

The full abstract domain is an extension of sized type schemas to
several predicate variables.
Each abstract element is a triple $\left\langle t, d, r
\right\rangle$ such that:
\begin{enumerate}
\vspace{-2mm}
\item $t$ is a set of $v \to (sized(\tau),c)$, where $v$ is a
  variable, $\tau$ its regular type and $c$ is its
  classification. Subgoal variables can be classified as
  \emph{output}, \emph{relevant}, or \emph{irrelevant}. Variables
  appearing in the clause body but not in the head are classified as
  \emph{clausal};
\item $d$ (the \emph{domain}) is a set of constraints over the
     relevant variables;
\item $r$ (the \emph{relations}) is a set of relations among bound variables. 
\end{enumerate}

For example, the final abstract elements corresponding to the clauses
of the \verb"listfact" example can be found below. The equations have
already been normalized into their simplest form,
and the variables refer to the predicate arguments in normal form.
$listfact$ refers implicitly to the solution of the joint equations:
it is the recurrence we need to solve.
In order to enhance readability, we have dropped the position element
$\langle ., 1 \rangle$ from $ln$.

\vspace{-2mm}
\begin{small}
$$\lambda'_1 = \left\langle
\begin{array}{c}
\left\{ L \to (ln^{(\alpha_1,\beta_1)}(n^{(\gamma_1,\delta_1)}), rel.),
       FL \to (ln^{(\alpha_2,\beta_2)}(n^{(\gamma_2,\delta_2)}), out.) \right\} \\
\{ \alpha_1 = 1, \beta_1 = 1 \},
\{ ln^{(\alpha_2,\beta_2)}(n^{(\gamma_2,\delta_2)}) \lessgtr ln^{(1,1)}(n^{nob}) \}
\end{array}
\right\rangle$$

$$\lambda'_{2} = \left\langle
\begin{array}{c}
\left\{ 
\begin{array}{c}
L \to (ln^{(\alpha_1,\beta_1)}(n^{(\gamma_1,\delta_1)}), rel.),
FL \to (ln^{(\alpha_2,\beta_2)}(n^{(\gamma_2,\delta_2)}), out.), \\
E \to (n^{(\gamma_3,\delta_3)}, cl.),
R \to (ln^{(\alpha_4,\beta_4)}(n^{(\gamma_4,\delta_4)}), cl.), \\
F \to (n^{(\gamma_5,\delta_5)}, cl.),
FR \to (ln^{(\alpha_6,\beta_6)}(n^{(\gamma_6,\delta_6)}), cl.)
\end{array}
\right\} \\
\{ \alpha_1 > 0, \beta_1 > 0 \}, \\
\left\{
\begin{array}{c}
%\dots \\
ln^{(\alpha_2, \beta_2)}(n^{(\gamma_2, \delta_2)}) 
 \lessgtr
 ln^{(\alpha' + 1, \beta' + 1)}(n^{(\min(\gamma_1 !, \gamma'), \max(\delta_1 !, \delta')}) \\
ln^{(\alpha', \beta')}(n^{(\gamma', \delta')}) 
 \lessgtr 
 listfact\left(ln^{(\alpha_1 - 1, \beta_1 - 1)}(n^{(\gamma_1, \delta_1)}) \right)
\end{array}
\right\}
\end{array}
\right\rangle$$
\vspace{-3mm}
\end{small}

%\vspace{-2mm}

\section{The Resources Abstract Domain}
\label{sec:resources}

We take advantage of the added power of sized types to develop a
better resource analysis which infers upper and lower bounds on the
amount of resources used by each predicate as a function of the sized
type schemas of the input arguments (which encode the sizes of the
terms and subterms appearing in such input arguments).
%, as output sizes were.
For this reason, the novel abstract domain for resource analysis that
we have developed is tightly integrated with the sized types abstract
domain.
% This guides as into developing a new resource analysis abstract domain
% integrated with the sized types one.
Following~\cite{resource-iclp07}, we account for two places where the
resource usage can be abstracted:
% modified:
\begin{itemize}

\item When entering a clause: some resources may be needed during
  unification of the call (subgoal) and the clause head, the
  preparation of entering that clause, and any work done when all the
  literals of the clause have been processed. This cost, dependent on
  the head $h$, is called \emph{head cost}, $\headcost(h)$.
  % and represented by $\beta$.

\item Before calling a literal $q$: some resources may be used to prepare
  a call to a body literal (e.g., constructing the actual
  arguments). The amount of these resources is known as
  \emph{literal cost} and is represented by $\litcost(q)$.
\end{itemize}

We first consider the case of estimating upper bounds on resource
usages. For simplicity, assume first
% also 
that we deal with predicates
having a behavior that is close to functional or imperative programs,
i.e., that are deterministic and do not fail. Then, we can bound the 
resource consumption of a clause $C \equiv p(\bar{x}) \imp
  q_1(\bar{x}_1), \dots, q_n(\bar{x}_n)$, denoted $r_{U,clause}$:
$$r_{U,clause}(C)
  \leq \headcost(p(\bar{x})) + \textstyle\sum_{i=1}^n
    \left( \litcost(q_i(\bar{x}_i)) + r_{U,pred}(q_i(\bar{x}_i)) \right)$$

%c

    As in sized type analysis, the sizes of some input arguments may
    be explicitly computed, or, otherwise, we express them by using a 
    generic expression, giving rise (in the case of recursive clauses)
    to a recurrence equation that we need to solve in order to find
    closed form resource usage functions.

    The resource usage of a predicate, $r_{U,pred}$, depending on its
    input data sizes, is obtained from the resource usage of the
    clauses defining it, by taking the maximum of the equation expressions that
    meet the constraints on the input data sizes (i.e., have
    the same domain).

    %% We only need to take care of taking the maximum when two equations
    %% apply to the same domain. 
    % Doing this we finish with the upper bound of the resource usage of a
    % predicate, $r_{U,pred}$.

In addition, we need to deal with two extra features of logic programming:
% However, in logic programming we have two extra features to take care of:
\begin{itemize}
\item We may execute a literal more than once on backtracking. To
  bound the number of times a literal is executed, we need to know the
  \emph{number of solutions} each literal (to its left) can
  generate. Using the information provided by cardinality analysis,
  the number of times a literal is
  executed is at most the product of the upper bound on the number of
  solutions, $s_U$, of all the previous literals in the clause. We get:

\vspace{-2mm}
\begin{small}
$$
\begin{array}{l}
r_{U,clause}\left( p(\bar{x}) \imp q_1(\bar{x}_1), \dots, q_n(\bar{x}_n) \right) \\
\quad  \leq \headcost(p(\bar{x})) + \sum_{i=1}^n 
    \left( \prod_{j = 1}^{i-1}s_{pred}(q_j(\bar{x}_j)) \right)
    \left( \litcost(q_i(\bar{x}_i)) + r_{U,pred}(q_i(\bar{x}_i)) \right)
\end{array}
$$
\end{small}
\vspace{-2mm}

%  We will look more in depth into bounding the number of solutions in
%  the following section.
\item Also, in logic programming more than one clause may unify with a
  given subgoal. In that case it is incorrect to take the maximum of
  the resource usages of each clause when setting up the recurrence
  equations (whereas this was valid in size analysis).
  A correct solution is to take the sum of every 
  set of equations with a common domain, but the bound becomes then
  very rough. Finer-grained possibilities can be considered by 
  using different \emph{aggregation} procedures per resource.
\end{itemize}

Lower bounds analysis is similar, but needs to take into account the
possibility of failure, which stops clause execution and forces
backtracking. Basically, no resource usage should be added beyond the
point where failure may happen.  For this reason, in our
implementation we use the non-failure analysis
already present in \ciaopp. Also, the aggregation of clauses with a
common domain must be different to that used in the upper bounds
case. The simplest solution is to just take the minimum of the
clauses. However, this again leads to very rough bounds. We will
discuss lower bound aggregation later.

%% \begin{itemize}
%% \item Basically, no resource usage should be added beyond the point
%%   where failure may happen. % This leads to the conclusion that we
%%   % need for some additional analysis to determine when a predicate will 
%%   % fail or not. In \ciaopp\ there is already a non-failure analysis,
%%   % which we use in our implementation of the abstract domain.
%%   For this reason, in our implementation of the abstract domain 
%%   we use the non-failure analysis already present in \ciaopp.

%% \item Aggregation of clauses with a common domain must be different
%%   to that used in the upper bounds case. The simplest solution is to just take
%%   the minimum of the clauses. However, this again leads to very rough
%%   bounds. We will discuss lower bounds aggregation later.
%% \end{itemize}

\paragraph{\textbf{Cardinality Analysis.}}
We have already discussed why cardinality analysis (which estimates
bounds on the number of solutions) is instrumental in resource
analysis of logic programs. We can consider the number of
solutions as another resource, but, due to its importance, we treat 
it separately.

An upper bound on the number of solutions of a single clause could be
gathered by multiplying the number of solutions of its body literals:
% all possible clauses:
$$s_{U,clause}\left( p(\bar{x}) \imp q_1(\bar{x}_1), \dots,
q_n(\bar{x}_n) \right) \leq \textstyle\prod_{i=1}^n
s_{U,pred}(q_i(\bar{x_i}))$$ For aggregation we need to add the
equations with a common domain, to get a recurrence equation
system. These equations will be solved later to get a closed form
function giving an upper bound on the number of solutions.

It is important to remark that many improvements can be added to
this simple cardinality analysis to make it more precise. Some of them
are discussed in~\cite{caslog}, like maintaining separate bounds for
the relation defined by the predicate and the number of solutions for
a particular input, or dealing with mutually exclusive clauses by
performing the $\max$ operation, instead of the addition operation
when aggregating. However, our focus here is the definition of an
abstract domain, and see whether a simple definition produces
comparable results for the resource usage analysis.

One of the improvements we decided to include is the use of the
% powerful 
determinacy analysis present in
\ciaopp~\cite{determinacy-ngc09}. If such analysis infers that a
predicate is deterministic, we can safely set the upper bound for the
number of solutions to 1.
%, avoiding the setting up of recurrence
%equations.

In the case of lower bounds, we need to know for each clause whether
it may fail or not. For that reason we use the non-failure analysis
already present in \ciaopp~\cite{nfplai-flops04}. In case of a
possible failure, the lower bound on cardinality is set to
0.

% \subsection{Aggregation}

% As we mentioned before, an important step when setting up the
% recurrence equations is the way in which we aggregate information
% becoming from different clauses. In relation to resources,
% traditionally the aggregation for upper bound has been performed in
% three phases:
% \begin{enumerate}
% \item Separate the clauses into mutually exclusive sets of them. We
%   know that for any input argument, it will only execute clauses in
%   one of these sets,
% \item For each mutually exclusive set, generate recurrence equations
%   by summing up all the equations from clauses that have a common
%   domain,
% \item Now we have a collection of systems of recurrences, and we need
%   to generate a final system to pass to the recurrence solver. We do
%   so by taking the maximum in each common domain.
% \end{enumerate}
% And a similar procedure can be developer for lower bounds.

% However, this approach restricts the set of resources that can be
% modeled in our framework. In our domain, resources can be endorsed
% with any kind of operation that, given the equations for all clauses
% and optionally some extra information, creates the set of equations to
% solve. This allows, for example, to model time in parallel programs as
% a resource, but where aggregation is never done by summing, but by
% taking the maximum in each case.

\paragraph{\textbf{The Abstract Elements.}}
% For CoRR
% Within the PLAI abstract interpretation framework~\cite{ai-jlp,inc-fixp-sas}
Within the PLAI abstract interpretation framework \\ \cite{ai-jlp,inc-fixp-sas}
an analysis is defined by the abstract elements involved in it
and a set of operations. 
We refer the reader to the~\ref{sec:absframe}
for an overview of the overall framework. In our case,
the abstract elements are derived from sized type analysis by adding some
extra components. In particular:
%, we include four new elements:
\begin{enumerate}
\item The \emph{current variable for solutions}, and \emph{current
    variable for each resource}.
\item A boolean element for telling whether we have already found a
  failing literal.
\item An abstract element from the non-failure domain.
\item An abstract element encoding information about determinacy.
\end{enumerate}
We will denote the abstract elements by
$\langle (s_L, s_U), v_{resources}, failed?, d, r, nf, det \rangle$
where $(s_L, s_U)$ are the lower and upper bound variables for the number
of solutions, $v_{resources}$ is a set of pairs $(r_L, r_U)$ giving
the lower and upper bound variables for each resource, $failed?$ is a
boolean element ({\tt true} or {\tt false}), $d$ and $r$ are
defined as in the sized type abstract domain, and $nf$ and $det$ 
can take values
% are summarized by the labels 
{\tt not\_fails}/{\tt fails} and {\tt
  non\_det}/{\tt is\_det} respectively,
as explained in \cite{determinacy-ngc09,nfplai-flops04}. 
\ref{sec:sem} gives some more details of the domain.

% In this analysis 
We assume that we are given the definition of a set
of resources, which are fixed throughout the whole analysis
process. 
% Sumarising a bit -PLG
We assume that for each resource $r$ we have:
% We have already mentioned three operations, but we need an extra one
% for having a complete algorithm. For each resource $r$ we have:
%\begin{itemize}
%\item
its head cost, $\headcost_r$, which takes a clause head as parameter;
%\item
its literal cost, $\litcost_r$, which takes a literal as parameter;
%\item
its aggregation procedure, $\Gamma_r$, which takes the equations
  for each of the clauses and creates a new set of recurrence
  equations from them; and
%\item
the default upper $\bot_{r,U}$ and lower $\bot_{r,L}$ bound on
      resource usage. 
%\end{itemize}

To better understand how the domain works, we will continue with the
analysis of \verb"listfact" that we started in the
previous section. We assume that the only resource to be analyzed is
the ``number of resolution steps,'' which uses the following parameters:

\vspace{-4mm}
$$\headcost = 1, \quad \litcost = 0, \quad \Gamma_r = +,
\quad (\bot_L, \bot_U) = (0,0)$$

\paragraph{\textbf{The $\sqsubseteq$, $\sqcup$ Operations and the $\bot$ Element.}}
We do not have a decidable definition for $\sqsubseteq$ or $\sqcup$,
because there is no general algorithm for checking the inclusion or
union of sets of integers defined by recurrence relations. Instead, for the inequation components we
just check whether one is a subset of another one,
up to variable renaming, or perform a syntactic union of the
inequations. The ordering is finished by taking the product order
with the non-failure and determinacy parts. This is enough for having a correct analysis.
For the bottom element, $\bot$, we first generate new variables for each of the resources
and the solution. Then, we add relations between them and the default
cost for each resource. For an unknown predicate, the number of
solutions should be $[0,
\infty)$
% Due to this same lack of information, we should 
and it may fail. 
%
% The components for non-failure and determinacy come from their
% corresponding abstract domains.
% \infty)$. Due to this same lack of information, we should assume that
% the predicate may have failed. 
% Commented out, redundant -PLG
% As mentioned before, the components for non-failure and determinacy
% come from the abstract domains for those analyses.
%
For example, the bottom element for the ``number of resolution steps'' resource will be:
$$\langle (s_L, s_U), \{ (n_L, n_U) \}, \texttt{true}, \emptyset, 
\{ (s_L, s_U) \lessgtr (0, \infty), (n_L, n_U) \lessgtr (0,0) \},
\texttt{fails}, \texttt{non\_det} \rangle$$
where {\tt fails} and {\tt non\_det} are the bottom elements of their respective domains.

\paragraph{\textbf{The $\lambda_{call}$ to $\beta_{entry}$ Operation.}}
In this operation we need to create the initial structures for
handling the bounds on the number of solutions and resources. This implies the generation
of fresh variables for each of them, and setting them to their initial
values. In the case of the number of solutions, the initial value is 1
(which is the number of solutions generated by a fact). For a resource
$r$, the initial value is exactly $\headcost_r$. We will name new fresh variables
by adding an integer subscript. For example,
$s_{L,1,1}$ will be the first fresh variable related to the \emph{l}ower bound on
\emph{s}olutions on \emph{first} clause.

The addition of constraints over sized types when the head arguments
are partially instantiated is inherited from the sized types domain.
Finally, for the $failed?$ component, we should start with value
\texttt{false}, as no literal has been executed yet, so it cannot
fail.

In the \verb"listfact" example, the entry substitutions are:

\vspace{-3mm}
\begin{small}
$$\beta_{entry,1} = 
\left\langle
\begin{array}{c}
(s_{L,1,1}, s_{U,1,1}), \{ (n_{L,1,1}, n_{U,1,1}) \}, \texttt{false},
\{ \alpha_1 = 0, \beta_1 = 0 \}, \\
\{ (s_{L,1,1}, s_{U,1,1}) \lessgtr (1,1), (n_{L,1,1}, n_{U,1,1}) \lessgtr (1,1) \},
\texttt{not\_fails}, \texttt{is\_det}
\end{array}
\right\rangle$$
$$\beta_{entry,2} = 
\left\langle
\begin{array}{c}
(s_{L,2,1}, s_{U,2,1}), \{ (n_{L,2,1}, n_{U,2,1}) \}, \texttt{false},
\{ \alpha_1 > 0, \beta_1 > 0 \}, \\
\{ (s_{L,2,1}, s_{U,2,1}) \lessgtr (1,1), (n_{L,2,1}, n_{U,2,1}) \lessgtr (1,1) \},
\texttt{not\_fails}, \texttt{is\_det}
\end{array}
\right\rangle$$
\end{small}
\vspace{-2mm}
%\vspace{-6mm}

\vspace{-1mm}
\paragraph{\textbf{The Extend Operation.}}
In the \emph{extend} operation we get both the current abstract substitution and the
substitution from the literal call. We need to update
several components of the abstract element. First of all, we need to
include a call to the function giving the number of solutions and
the resource usage from the called literal.

Afterwards, we need to generate new variables for the number of solutions and
resources, which will hold the bounds for the clause up to that
point. New relations must be added to the abstract element to give a
value to those new variables:
\begin{itemize}
\item For the number of solutions, let $s_{U,c}$ be the new upper bound
  variable, $s_{U,p}$ the previous variable defining an upper bound on
  the number of solutions, and $s_{U,\lambda}$ an upper bound on the number of
  solutions for the subgoal. Then we need to include a constraint:
  $s_{U,c} \leq s_{U,p} \times s_{U,\lambda}$.

  In the case of lower bound analysis, there are two phases. First of all, we
  check whether the called literal can fail, looking at the output of
  the non-failure analysis. If it is possible for it to fail, we update
  the $failed?$ component of the abstract element to {\tt true}. If
  after this checking the $failed?$ component is still {\tt false} (meaning
  that neither this literal nor any of the previous ones may fail) we
  include a relation similar to the one for the upper bound case: $s_{L,c} \geq
  s_{L,p} \times s_{L,\lambda}$. Otherwise, we include the relation
  $s_{L,c} \geq 0$, because failing predicates produce no solutions.

\item The approach for resources is similar. Let $r_{U,c}$ be the new
  upper bound variable, $r_{U,p}$ the previous variable defining an upper
  bound on that resource and $r_{U,\lambda}$ an upper bound on
  resources from the analysis of the literal. The relation added in
  this case is $r_{U,c} \leq r_{U,p} + s_{U,p} \times \left( \litcost +
    r_{U,\lambda} \right)$.

  For lower bounds, we have already updated the $failed?$ component,
  so we only have to work in consequence. If the component is still
  {\tt false}, we add a new relation similar to the one for upper
  bounds. If it is {\tt true}, it means that failure may happen at
  some point, so we do not have to add that resource any more. Thus
  the relation to be included is $r_{L,c} \geq r_{L,p}.$
\end{itemize}
In our example, consider the extension of \verb"listfact" after
performing the analysis of the \verb"fact" literal, whose resource
components of the abstract element will be:

\vspace{-2mm}
\begin{small}
$$\left\langle
\begin{array}{c}
(s_{L}, s_{U}), \{ (n_{L}, n_{U}) \}, \texttt{false}, \{ \alpha, \beta \geq 0 \} \\
\{ (s_{L}, s_{U}) \lessgtr (1,1), (n_{L}, n_{U}) \lessgtr (\alpha,\beta) \}, 
\texttt{not\_fails}, \texttt{is\_det}
\end{array}
\right\rangle$$
\end{small}

\noindent This literal is known not to fail, so we do not change
the value of $failed?$ in our abstract element for the
second clause. That means that it is still {\tt false}, so we add
complete calls:

\vspace{-2mm}
\begin{small}
$$\beta_{entry,2} = 
\left\langle
\begin{array}{c}
(s_{L,2,2}, s_{U,2,2}), \{ (n_{L,2,2}, n_{U,2,2}) \}, \texttt{false}, \{ \dots \} \\
\left\{
\begin{array}{c}
 \dots, \\
 (s_{L,2,2}, s_{U,2,2}) \lessgtr (1 \times s_{L,2,1}, 1 \times s_{U,2,1}), \\
(n_{L,2,2}, n_{U,2,2}) \lessgtr (\gamma_1 + n_{L,2,1}, \delta_1 + n_{U,2,1})
\end{array} \right\}, \\
\texttt{not\_fails}, \texttt{is\_det}
\end{array}
\right\rangle$$
\end{small}
\vspace{-3mm}
%\vspace{-6mm}

\vspace{-2mm}
\paragraph{\textbf{The $\beta_{exit}$ to $\lambda'$ Operation.}}
After all the extend operations, the variables appearing in
the number of solutions and resources positions will hold the correct
value for their properties. As we did with sized
types, we follow now a normalization step, based on ~\cite{caslog}:
replace each variable appearing in an expression with its
definition in terms of other variables, in reverse topological order.
Following this process, we should reach the variables in
the sized types of the input parameters in the head.

Going back to \verb"listfact", the final substitutions are as follows.
$s'_L, s'_U, n'_L$ and $n'_U$ refer to number of solutions and resolution
steps from the recursive call to {\tt listfact}.

\vspace{-2mm}
\begin{small}
$$\lambda'_{1} = 
\left\langle
\begin{array}{c}
(s_{L,1,1}, s_{U,1,1}), \{ (n_{L,1,1}, n_{U,1,1}) \}, \texttt{false},
\{ \alpha_1 = 0, \beta_1 = 0 \}, \\
\{ (s_{L,1,1}, s_{U,1,1}) \lessgtr (1,1), (n_{L,1,1}, n_{U,1,1}) \lessgtr (1,1) \},
\texttt{not\_fails}, \texttt{is\_det}
\end{array}
\right\rangle$$
$$\lambda'_{entry,2} = 
\left\langle
\begin{array}{c}
(s_{L,2,3}, s_{U,2,3}), \{ (n_{L,2,3}, n_{U,2,3}) \}, \texttt{false},
\{ \alpha_1 > 0, \beta_1 > 0 \}, \\
\left\{
\begin{array}{c}
s_{L,2,3} \geq 1 \times s'_{L}(ln^{(\alpha_1 - 1,\beta_1 - 1)}(n^{(\gamma_1,\delta_1)})), \\
s_{U,2,3} \leq 1 \times s'_{U}(ln^{(\alpha_1 - 1,\beta_1 - 1)}(n^{(\gamma_1,\delta_1)})), \\
n_{L,2,3} \geq \gamma_1 + n'_{L}(ln^{(\alpha_1 - 1,\beta_1 - 1)}(n^{(\gamma_1,\delta_1)})), \\
n_{U,2,3} \leq \delta_1 + n'_{U}(ln^{(\alpha_1 - 1,\beta_1 - 1)}(n^{(\gamma_1,\delta_1)}))
\end{array}
\right\}, \\
\texttt{not\_fails}, \texttt{is\_det}
\end{array}
\right\rangle$$
\vspace{-3mm}
\end{small}

\vspace{-2mm}
\paragraph{\textbf{The Widening Operator $\nabla$ and Closed Forms.}}
As mentioned before, in contrast to previous cost analyses, at this
point we bring in the possibility of different aggregation
operators. Thus, when we have the equations, we need to pass them to
each of the corresponding $\Gamma_r$ per each resource $r$ to get the
final equations.

This process can be further refined in the case of solution analysis,
using the information from the non-failure and determinacy analyses.
If the final output of the non-failure analysis is {\tt fails}, we
know that the only correct lower bound is 0. So we can just assign the
relation $s_L \geq 0$ without further relations.
Conversely, if the final output of the determinacy analysis is {\tt
  is\_det}, we can safely set the relation $s_U \leq 1$, because at
most one solution will be produced in each case. Furthermore, we can
refine the lower bound on the number of solutions with the minimum
between the current bound and 1.
% \begin{itemize}
% \item If the final output of the non-failure analysis is {\tt fails},
%   we know that the only correct lower bound is 0. So we can just
%   assign the relation $s_L \geq 0$ without further recurrence relation setting.
% \item If the final output of the determinacy analysis is {\tt
%     is\_det}, we can safely set the relation $s_U \leq 1$, because at
%   most one 
%   solution will be produced in each case. Furthermore, we can refine
%   the lower bound on the number of solutions with the minimum
%   between the current bound and 1.
% \end{itemize}

In the example analyzed above there was an implicit assumption while
setting up the relations: that the recursive call in the body of
\verb"listfact" refers to the same predicate call, so we can set up a
recurrence. This fact is implicitly assumed in Hindley-Milner
type systems.
%, where each expression and function receives only one type.
But in logic programming it is usual for a predicate to be
called with different patterns (for example,
modes). Fortunately, the \ciaopp\ framework allows multivariance
(support for different call patterns of the same predicate).
For the analysis to handle it, we cannot just add calls with
the bare name of the predicate, because it will conflate all the
versions. The solution is to add a new component to the
abstract element: a random name given to the specific instance of the
predicate, and generated in the $\lambda_{call}$
to $\beta_{entry}$. In the widening step, all different versions
of the same predicate are conflated.

Even though the analysis works with relations, these are not as useful
as functions defined without recursion or calls to other
functions. First of all, developers will get a better idea of the
sizes presented in such a closed form. Second, functions are
amenable to comparison as outlined
in~\cite{resource-verif-iclp2010}, which is essential
in verification. There are several
packages able to get bounds for recurrence equations:
computer algebra systems, such as Mathematica
(which has been used in our experiments) 
% integrated to get a fully automated analysis)
%(the one used in our experiments)
or Maxima; and specialized solvers such as
PURRS~\cite{BagnaraPZZ05} or PUBS
\cite{conf-vmcai-AlbertGM11}. In our implementation we apply
this overapproximation operator after each widening.  For our
example, the final abstract substitution is:

\vspace{-3mm}
\begin{small}
$$\lambda'_1 \nabla \lambda'_2 =
\left\langle
\begin{array}{c}
(s_{L}, s_{U}), \{ (n_{L}, n_{U}) \}, \texttt{false}, \{ \alpha_1, \beta_1 \geq 0 \}, \\
\left\{ (s_L, s_U) \lessgtr (1,1), (n_L, n_U) \lessgtr (\alpha_1 \gamma_1, \beta_1 \delta_1) \right\},
\texttt{not\_fails}, \texttt{is\_det}
\end{array}
\right\rangle$$
\vspace{-3mm}
\end{small}

% \noindent The system presents the output to the user in the form of
% Ciao assertions:
% \begin{footnotesize}
% \begin{lstlisting}
% :- true pred listfact(_A,_B) : ( list(_A,num), var(_B) )
%      => ( list(_A,num), list(_B,num),
%           rsize(_A,'basic_props:list'(_C,_D,num(_E,_F))),
%           rsize(_B,'basic_props:list'(_C,_D,
%                 num('Factorial'(_E),'Factorial'(_F)))) )
%       + ( cardinality(1,1), resource(steps,_C*_E,_D*_F) ).
% \end{lstlisting}
% \end{footnotesize}

% An $\uparrow$ operator will try to replace relations with a closed
% form bound. We can see this operator as overapproximating an abstract
% element, $x \sqsubseteq \, \uparrow x$.  There are several software
% packages with these capabilities: computer algebra systems, such as
% Mathematica (the one used in our experiments) or Maxima; and
% specialized solvers such as PURRS~\cite{BagnaraPZZ05} or
% PUBS~\cite{conf-vmcai-AlbertGM11}. In our implementation we
% apply 
% this overapproximation operator after each widening step. More refined
% strategies to decide when to solve could be developer, but we found
% that this simple heuristic works well enough in practice.

%\vspace{-7mm}

\begin{table}[t]
\caption{Experimental results.}\label{expresults}
%\vspace{-2mm}
% \vspace{0.5\baselineskip}
\begin{minipage}{\textwidth}
\begin{center}
%\begin{small}
%\begin{footnotesize}
\begin{tabular}{|l|ccc|ccccc|cc|}
\hline
\hline
\textbf{\emph{Program}}
& \multicolumn{3}{c|}{\textbf{\emph{Resource A. (LB)}}}
& \multicolumn{5}{c|}{\textbf{\emph{Resource A. (UB)}}}
% & \multicolumn{3}{c|}{\textbf{\emph{Resource Analysis (LB)}}}
% & \multicolumn{5}{c|}{\textbf{\emph{Resource Analysis (UB)}}}
& \multicolumn{2}{c|}{\textbf{\emph{A. Times (s)}}}
\\
& \emph{New} & \multicolumn{2}{c|}{\emph{Prev.}} 
& \emph{New} & \multicolumn{2}{c}{\emph{Prev.}} 
& \multicolumn{2}{c|}{\emph{RAML}} 
& \emph{New} & \emph{Prev.} \\ 
\hline \hline
\texttt{append}
& $\alpha$
& $\alpha$
& = 
& $\beta$
& $\beta$
& =
& $\beta$
& = 
& 0.999 
& 0.530 \\
\texttt{appendAll2}
& $a_1a_2a_3$
& $a_1$
& +
& $b_1b_2b_3$
& $\infty$
& +
& $b_1b_2b_3$
& = 
& 2.408    
& 0.668   \\
\texttt{coupled}
& $\mu$
& $0$
& +
& $\nu$
& $\infty$
& +
& $\nu$
& = 
& 1.365    
& 0.644  \\
\texttt{dyade}
& $\alpha_1\alpha_2$
& $\alpha_1\alpha_2$
& = 
& $\beta_1\beta_2$
& $\beta_1\beta_2$
& =
& $\beta_1\beta_2$
& = 
& 1.658    
& 0.620  \\
\texttt{erathos}
& $\alpha$
& $\alpha$
& = 
& $\beta^2$
& $\beta^2$
& =
& $\beta^2$
& = 
& 2.251    
& 0.772 \\
\texttt{fib}
& $\phi^\mu$
& $\phi^\mu$
& = 
& $\phi^\nu$
& $\phi^\nu$
& =
& infeasible
& + 
& 1.064    
& 0.671 \\
\texttt{hanoi}
& $1$
& $0$
& +
& $2^\nu$
& $\infty$
& +
& infeasible
& + 
& 0.819     
& 0.603 \\
\texttt{isort}
& $\alpha^2$
& $\alpha^2$
& = 
& $\beta^2$
& $\beta^2$
& =
& $\beta^2$
& = 
& 1.675     
& 0.617   \\
\texttt{isortlist}
& $a_1^2$
& $a_1^2$
& = 
& $b_1^2b_2$
& $\infty$
& +
& $b_1^2b_2$
& = 
& 2.546    
& 0.669    \\
\texttt{listfact}
& $\alpha\gamma$
& $\alpha$
& +
& $\beta\delta$
& $\infty$
& +
& unknown
& ? 
& 1.387    
& 0.644      \\
\texttt{listnum}
& $\mu$
& $\mu$
& = 
& $\nu$
& $\nu$
& =
& unknown
& ? 
& 1.189    
& 0.581    \\
\texttt{minsort}
& $\alpha^2$
& $\alpha$
& +
& $\beta^2$
& $\beta^2$
& =
& $\beta^2$
& = 
& 1.938    
& 0.671   \\
\texttt{nub}
& $a_1$
& $a_1$
& = 
& $b_1^2b_2$
& $\infty$
& +
& $b_1^2b_2$
& = 
& 3.614    
& 0.910 \\
\texttt{partition}
& $\alpha$
& $\alpha$
& =
& $\beta$
& $\beta$
& =
& $\beta$
& = 
& 1.698    
& 0.647 \\
\texttt{zip3}
& $\min(\alpha_i)$
& $0$
& +
& $\min(\beta_i)$
& $\infty$
& +
& $\beta_3$
& + 
& 2.484    
& 0.570 \\
\hline
\end{tabular}
%\end{small}
\end{center}
\end{minipage}
\vspace{-2\baselineskip}
% \vspace{-0.7cm}
\end{table}

\vspace{-0.3cm}

\section{Experimental Results}
\label{sec:results}

We have constructed a prototype implementation in \ciao{} by defining
the abstract operations for sized type and resource analysis that we
have described and plugging them into \ciaopp's PLAI.
Our objective is to assess the gains in precision in
resource analysis.

Table~\ref{expresults} shows the results of the comparison between the
new lower (\textbf{\emph{LB}}) and upper bound (\textbf{\emph{UB}})
resource analyses implemented in \ciaopp{}, which also use the new
size analysis (columns \emph{New}), and the previous resource analyses
in \ciaopp{}~\cite{caslog,low-bounds-ilps97,resource-iclp07} (columns
\emph{Prev.}). We also compare (for upper bounds) with
\emph{RAML} ~\cite{DBLP:journals/toplas/0002AH12}.
Although the new resource analysis and the previous one infer concrete
resource usage bound functions, 
for the sake of conciseness and to make
the comparison with RAML meaningful, 
Table~\ref{expresults} only shows
the complexity orders of such functions, e.g., if the analysis infers
the resource usage bound function $\Phi$, and $\Phi \in
\Theta(\Psi)$, Table~\ref{expresults} shows $\Psi$. The parameters
of such functions are (lower or upper) bounds on input data sizes. The
symbols used to name such parameters have been chosen assuming that
lists of numbers $L_i$ have size
$ln^{(\alpha_i,\beta_i)}(n^{(\gamma_i,\delta_i)})$, lists of lists of
lists of numbers have size
$llln^{(a_1,b_1)}(lln^{(a_2,b_2)}(ln^{(a_3,b_3)}(n^{(a_4,b_4)})))$, and
numbers have size $n^{(\mu,\nu)}$. The calling modes
are the usual ones with the last argument as output.

Table~\ref{expresults}
includes columns with symbols summarizing whether the new
\ciaopp\ resource analysis improves on the previous one and
\emph{RAML}'s: $+$ (resp. $-$) indicates more (resp.\ less)
precise bounds, and $=$ the same.  % We can see that t
The new resource 
% size 
analysis improves on \ciaopp's previous analysis.
Moreover, RAML can only infer polynomial costs, while
our approach is able to infer
other types of functions, as shown for the divide-and-conquer
benchmarks \texttt{hanoi} and \texttt{fib}, which represent a
common class of programs. For predicates with polynomial cost, we
get equal or better results than RAML.
%
% Efficiency

The last two columns show the times (in seconds) required by both 
lower and upper bound % resource 
analysis together for the new resource
analysis, and for the previous resource analysis in \ciaopp{}
(\ciao/\ciaopp~version 1.15-2124-ga588643, on an Intel Core i7 2.4
GHz, 8 GB 1333 MHz DDR3 memory, running MAC OS X Lion 10.7.5).
% , and averaging several runs, eliminating the best and worst values),
% 
These times include also the auxiliary non-determinism and failure analyses.
The resulting times are encouraging, despite the currently
relatively inefficient implementation of the interface
with the Mathematica system which is used for solving recurrence
equations. 

\vspace{-2mm}
\section{Related Work}
\label{sec:related}

Several other analyses for resources have been proposed in the
literature. Some of them just focus on one particular resource
(usually execution or heap consumption), but it seems clear that 
they could be generalized.
% without further
%problems.
% For CoRR
% We already mentioned RAML~\cite{DBLP:journals/toplas/0002AH12} in 
We already mentioned RAML \\ \cite{DBLP:journals/toplas/0002AH12} in 
Section~\ref{sec:results}. Their approach differs from ours in the
theoretical 
framework being used: RAML uses a type and effect system, whereas we
use abstract interpretation. Another difference is
the use of polynomials in RAML, which allows a complete
method of resolution but limits the type of closed forms that can be
analyzed. In contrast, we use recurrence equations, which have no complete
decision procedure, but encompass a much larger class of
functions. Type systems are also used to guide inference in
\cite{grobauer01cost} and~\cite{igarashi02resource}.
In~\cite{nielson02automatic}, the authors use sparsity information to
infer asymptotic complexities, instead of recurrences.
\cite{DBLP:conf/ppdp/GieslSSEF12} uses symbolic evaluation graphs
to derive termination and complexity properties.
The recurrence equation approach was proposed originally by
Wegbreit~\cite{Wegbreit75}.  Similarly to \ciaopp's previous analysis,
the approach of~\cite{conf-vmcai-AlbertGM11} applies the recurrence
equation method directly (i.e., not within an abstract interpretation
framework).  \cite{Rosendahl89} shows a complexity analysis based on
abstract interpretation over a step-counting version of functional
programs, but which does not generate closed forms.
Types with embedded size information have also been proposed 
by~\cite{DBLP:conf/ifl/VasconcelosH03} for functional programs. 
Our sized type % includes some enhancements to deal with 
analysis is based on regular types and abstract interpretation,
% in logic programs, and incorporates solutions to deal with the
% additional features of logic programming such as non-determinism and
% backtracking.  in logic programs,
and deals with the 
% additional 
logic programming features such as unification, non-determinism, and
backtracking.

\vspace{-2mm}
\section{Conclusions}

We have presented a new formulation of resource analysis as a domain
within abstract interpretation and which uses as input information the
sized types that we developed in~\cite{sized-types-iclp2013}.  Our
approach overcomes important limitations of existing resource analyses
and enhances their precision. It also benefits from an easier
implementation and integration within an abstract interpretation
framework such as PLAI/\ciaopp, which brings in useful features such
as \emph{multivariance} for free. Finally, the results of our
experimental assessment regarding accuracy and efficiency are quite
encouraging.

% offers benefits both in the quality of the bounds inferred by the
% analysis, and in the ease of implementation and integration within a
% framework such as PLAI/\ciaopp.

%% In the future, we would like to study the generalization of this
%% % resource analysis 
%% framework to
%% %allow the analysis of resources with slightly
%% different behaviors
%% regarding aggregation. For example, when running tasks in parallel,
%% the total time is basically the maximum of both tasks, but memory
%% usage is bounded by the sum of them.
%% % 
%% Another future direction is the use of more ancillary analyses
%% % integration of more of the analyses
%% %present in the \ciaopp\ analysis system, 
%% %in order
%% to obtain more precise
%% results. 
%% % Commented out -PLG
%% %% Also, since we use sized types as a basis, any new research that
%% %% improves such analysis will directly benefit the resource analysis.
%% %
%% %One of the first
%% %enhancements we are planning includes taking benefit from the variable
%% %sharing analyses for creating further constraints over sized types and
%% %for giving sharper bounds for solution analysis, and using 
%% Finally, another planned enhancement is the use of mutual exclusion
%% analysis (already present in \ciaopp) to aggregate recurrence
%% equations in a better way.

% \newpage

\bibliographystyle{acmtrans}

%% \begin{small}
%% 
%% \bibliography{../../../bibtex/clip/clip,../../../bibtex/clip/general}

\begin{thebibliography}{}

\bibitem[\protect\citeauthoryear{Albert, Genaim, and Masud}{Albert
  et~al\mbox{.}}{2011}]{conf-vmcai-AlbertGM11}
{\sc Albert, E.}, {\sc Genaim, S.}, {\sc and} {\sc Masud, A.~N.} 2011.
\newblock {M}ore {P}recise yet {W}idely {A}pplicable {C}ost {A}nalysis.
\newblock In {\em 12th Verification, Model Checking, and Abstract
  Interpretation (VMCAI'11)}, {R.~Jhala} {and} {D.~Schmidt}, Eds. Lecture Notes
  in Computer Science, vol. 6538. Springer Verlag, 38--53.

\bibitem[\protect\citeauthoryear{Bagnara, Pescetti, Zaccagnini, and
  Zaffanella}{Bagnara et~al\mbox{.}}{2005}]{BagnaraPZZ05}
{\sc Bagnara, R.}, {\sc Pescetti, A.}, {\sc Zaccagnini, A.}, {\sc and} {\sc
  Zaffanella, E.} 2005.
\newblock {PURRS}: Towards {C}omputer {A}lgebra {S}upport for {F}ully
  {A}utomatic {W}orst-{C}ase {C}omplexity {A}nalysis.
\newblock Tech. rep.
\newblock {\tt arXiv:cs/0512056} available from \url{http://arxiv.org/}.

\bibitem[\protect\citeauthoryear{Bruynooghe}{Bruynooghe}{1991}]{DBLP:journals/jlp/Bruynooghe91}
{\sc Bruynooghe, M.} 1991.
\newblock A practical framework for the abstract interpretation of logic
  programs.
\newblock {\em J. Log. Program.\/}~{\em 10,\/}~2, 91--124.

\bibitem[\protect\citeauthoryear{Bueno, L\'{o}pez-Garc\'{\i}a, and
  Hermenegildo}{Bueno et~al\mbox{.}}{2004}]{nfplai-flops04}
{\sc Bueno, F.}, {\sc L\'{o}pez-Garc\'{\i}a, P.}, {\sc and} {\sc Hermenegildo,
  M.} 2004.
\newblock {M}ultivariant {N}on-{F}ailure {A}nalysis via {S}tandard {A}bstract
  {I}nterpretation.
\newblock In {\em 7th International Symposium on Functional and Logic
  Programming (FLOPS 2004)}. Number 2998 in LNCS. Springer-Verlag, Heidelberg,
  Germany, 100--116.

\bibitem[\protect\citeauthoryear{Cousot and Cousot}{Cousot and
  Cousot}{1992}]{cousots-lp}
{\sc Cousot, P.} {\sc and} {\sc Cousot, R.} 1992.
\newblock {A}bstract {I}nterpretation and {A}pplications to {L}ogic {P}rograms.
\newblock {\em Journal of Logic Programming\/}~{\em 13,\/}~2-3, 103--179.

\bibitem[\protect\citeauthoryear{Dart and Zobel}{Dart and
  Zobel}{1992}]{Dart-Zobel}
{\sc Dart, P.} {\sc and} {\sc Zobel, J.} 1992.
\newblock {A} {R}egular {T}ype {L}anguage for {L}ogic {P}rograms.
\newblock In {\em {T}ypes in {L}ogic {P}rogramming}. MIT Press, 157--187.

\bibitem[\protect\citeauthoryear{Debray and Lin}{Debray and Lin}{1993}]{caslog}
{\sc Debray, S.~K.} {\sc and} {\sc Lin, N.~W.} 1993.
\newblock {Cost Analysis of Logic Programs}.
\newblock {\em {ACM} Transactions on Programming Languages and Systems\/}~{\em
  15,\/}~5 (November), 826--875.

\bibitem[\protect\citeauthoryear{Debray, Lin, and Hermenegildo}{Debray
  et~al\mbox{.}}{1990}]{granularity}
{\sc Debray, S.~K.}, {\sc Lin, N.-W.}, {\sc and} {\sc Hermenegildo, M.} 1990.
\newblock {T}ask {G}ranularity {A}nalysis in {L}ogic {P}rograms.
\newblock In {\em Proc. of the 1990 {ACM} Conf. on Programming Language Design
  and Implementation}. {ACM} Press, 174--188.

\bibitem[\protect\citeauthoryear{Debray, L\'{o}pez-Garc\'{\i}a, Hermenegildo,
  and Lin}{Debray et~al\mbox{.}}{1997}]{low-bounds-ilps97}
{\sc Debray, S.~K.}, {\sc L\'{o}pez-Garc\'{\i}a, P.}, {\sc Hermenegildo, M.},
  {\sc and} {\sc Lin, N.-W.} 1997.
\newblock {L}ower {B}ound {C}ost {E}stimation for {L}ogic {P}rograms.
\newblock In {\em 1997 International Logic Programming Symposium}. MIT Press,
  Cambridge, MA, 291--305.

\bibitem[\protect\citeauthoryear{Giesl, Str{\"o}der, Schneider-Kamp, Emmes, and
  Fuhs}{Giesl et~al\mbox{.}}{2012}]{DBLP:conf/ppdp/GieslSSEF12}
{\sc Giesl, J.}, {\sc Str{\"o}der, T.}, {\sc Schneider-Kamp, P.}, {\sc Emmes,
  F.}, {\sc and} {\sc Fuhs, C.} 2012.
\newblock Symbolic evaluation graphs and term rewriting: a general methodology
  for analyzing logic programs.
\newblock In {\em PPDP}. ACM, 1--12.

\bibitem[\protect\citeauthoryear{Grobauer}{Grobauer}{2001}]{grobauer01cost}
{\sc Grobauer, B.} 2001.
\newblock Cost recurrences for {DML} programs.
\newblock In {\em International Conference on Functional Programming}.
  253--264.

\bibitem[\protect\citeauthoryear{Hermenegildo, Bueno, Carro, L\'{o}pez, Mera,
  Morales, and Puebla}{Hermenegildo
  et~al\mbox{.}}{2012}]{hermenegildo11:ciao-design-tplp}
{\sc Hermenegildo, M.~V.}, {\sc Bueno, F.}, {\sc Carro, M.}, {\sc L\'{o}pez,
  P.}, {\sc Mera, E.}, {\sc Morales, J.}, {\sc and} {\sc Puebla, G.} 2012.
\newblock {A}n {O}verview of {C}iao and its {D}esign {P}hilosophy.
\newblock {\em Theory and Practice of Logic Programming\/}~{\em 12,\/}~1--2
  (January), 219--252.
\newblock http://arxiv.org/abs/1102.5497.

\bibitem[\protect\citeauthoryear{Hoffmann, Aehlig, and Hofmann}{Hoffmann
  et~al\mbox{.}}{2012}]{DBLP:journals/toplas/0002AH12}
{\sc Hoffmann, J.}, {\sc Aehlig, K.}, {\sc and} {\sc Hofmann, M.} 2012.
\newblock Multivariate amortized resource analysis.
\newblock {\em ACM Trans. Program. Lang. Syst.\/}~{\em 34,\/}~3, 14.

\bibitem[\protect\citeauthoryear{Igarashi and Kobayashi}{Igarashi and
  Kobayashi}{2002}]{igarashi02resource}
{\sc Igarashi, A.} {\sc and} {\sc Kobayashi, N.} 2002.
\newblock Resource usage analysis.
\newblock In {\em Symposium on Principles of Programming Languages}. 331--342.

\bibitem[\protect\citeauthoryear{L\'{o}pez-Garc\'{\i}a, Bueno, and
  Hermenegildo}{L\'{o}pez-Garc\'{\i}a et~al\mbox{.}}{2010}]{determinacy-ngc09}
{\sc L\'{o}pez-Garc\'{\i}a, P.}, {\sc Bueno, F.}, {\sc and} {\sc Hermenegildo,
  M.} 2010.
\newblock {A}utomatic {I}nference of {D}eterminacy and {M}utual {E}xclusion for
  {L}ogic {P}rograms {U}sing {M}ode and {T}ype {I}nformation.
\newblock {\em New Generation Computing\/}~{\em 28,\/}~2, 117--206.

\bibitem[\protect\citeauthoryear{L\'{o}pez-Garc\'{\i}a, Darmawan, and
  Bueno}{L\'{o}pez-Garc\'{\i}a et~al\mbox{.}}{2010}]{resource-verif-iclp2010}
{\sc L\'{o}pez-Garc\'{\i}a, P.}, {\sc Darmawan, L.}, {\sc and} {\sc Bueno, F.}
  2010.
\newblock {A} {F}ramework for {V}erification and {D}ebugging of {R}esource
  {U}sage {P}roperties.
\newblock In {\em {T}echnical {C}ommunications of the 26th {I}nt'l.
  {C}onference on {L}ogic {P}rogramming (ICLP'10)}, {M.~Hermenegildo} {and}
  {T.~Schaub}, Eds. Leibniz International Proceedings in Informatics (LIPIcs),
  vol.~7. Schloss Dagstuhl--Leibniz-Zentrum fuer Informatik, Dagstuhl, Germany,
  104--113.

\bibitem[\protect\citeauthoryear{Muthukumar and Hermenegildo}{Muthukumar and
  Hermenegildo}{1989}]{abs-int-naclp89}
{\sc Muthukumar, K.} {\sc and} {\sc Hermenegildo, M.} 1989.
\newblock {D}etermination of {V}ariable {D}ependence {I}nformation at
  {C}ompile-{T}ime {T}hrough {A}bstract {I}nterpretation.
\newblock In {\em 1989 North American Conference on Logic Programming}. {MIT}
  Press, 166--189.

\bibitem[\protect\citeauthoryear{Muthukumar and Hermenegildo}{Muthukumar and
  Hermenegildo}{1992}]{ai-jlp}
{\sc Muthukumar, K.} {\sc and} {\sc Hermenegildo, M.} 1992.
\newblock {C}ompile-time {D}erivation of {V}ariable {D}ependency {U}sing
  {A}bstract {I}nterpretation.
\newblock {\em Journal of Logic Programming\/}~{\em 13,\/}~2/3 (July),
  315--347.

\bibitem[\protect\citeauthoryear{Navas, Mera, L\'{o}pez-Garc\'{\i}a, and
  Hermenegildo}{Navas et~al\mbox{.}}{2007}]{resource-iclp07}
{\sc Navas, J.}, {\sc Mera, E.}, {\sc L\'{o}pez-Garc\'{\i}a, P.}, {\sc and}
  {\sc Hermenegildo, M.} 2007.
\newblock {U}ser-{D}efinable {R}esource {B}ounds {A}nalysis for {L}ogic
  {P}rograms.
\newblock In {\em 23rd International Conference on Logic Programming
  (ICLP'07)}. Lecture Notes in Computer Science, vol. 4670. Springer.

\bibitem[\protect\citeauthoryear{Nielson, Nielson, and Seidl}{Nielson
  et~al\mbox{.}}{2002}]{nielson02automatic}
{\sc Nielson, F.}, {\sc Nielson, H.~R.}, {\sc and} {\sc Seidl, H.} 2002.
\newblock Automatic complexity analysis.
\newblock In {\em European Symposium on Programming}. 243--261.

\bibitem[\protect\citeauthoryear{Puebla and Hermenegildo}{Puebla and
  Hermenegildo}{1996}]{inc-fixp-sas}
{\sc Puebla, G.} {\sc and} {\sc Hermenegildo, M.} 1996.
\newblock {O}ptimized {A}lgorithms for the {I}ncremental {A}nalysis of {L}ogic
  {P}rograms.
\newblock In {\em International Static Analysis Symposium (SAS 1996)}. Number
  1145 in {LNCS}. Springer-Verlag, 270--284.

\bibitem[\protect\citeauthoryear{Rosendahl}{Rosendahl}{1989}]{Rosendahl89}
{\sc Rosendahl, M.} 1989.
\newblock {A}utomatic {C}omplexity {A}nalysis.
\newblock In {\em 4th ACM {C}onference on {F}unctional {P}rogramming
  {L}anguages and {C}omputer {A}rchitecture (FPCA'89)}. ACM Press.

\bibitem[\protect\citeauthoryear{Serrano, Lopez-Garcia, Bueno, and
  Hermenegildo}{Serrano et~al\mbox{.}}{2013}]{sized-types-iclp2013}
{\sc Serrano, A.}, {\sc Lopez-Garcia, P.}, {\sc Bueno, F.}, {\sc and} {\sc
  Hermenegildo, M.} 2013.
\newblock {S}ized {T}ype {A}nalysis for {L}ogic {P}rograms (technical
  communication).
\newblock In {\em Theory and Practice of Logic Programming, 29th Int'l.
  Conference on Logic Programming (ICLP'13) Special Issue, On-line Supplement},
  {T.~Swift} {and} {E.~Lamma}, Eds. Vol.~13. Cambridge U. Press, 1--14.

\bibitem[\protect\citeauthoryear{Vasconcelos and Hammond}{Vasconcelos and
  Hammond}{2003}]{DBLP:conf/ifl/VasconcelosH03}
{\sc Vasconcelos, P.~B.} {\sc and} {\sc Hammond, K.} 2003.
\newblock Inferring cost equations for recursive, polymorphic and higher-order
  functional programs.
\newblock In {\em IFL}, {P.~W. Trinder}, {G.~Michaelson}, {and} {R.~Pena}, Eds.
  Lecture Notes in Computer Science, vol. 3145. Springer, 86--101.

\bibitem[\protect\citeauthoryear{Vaucheret and Bueno}{Vaucheret and
  Bueno}{2002}]{eterms-sas02}
{\sc Vaucheret, C.} {\sc and} {\sc Bueno, F.} 2002.
\newblock {M}ore {P}recise yet {E}fficient {T}ype {I}nference for {L}ogic
  {P}rograms.
\newblock In {\em International Static Analysis Symposium}. Lecture Notes in
  Computer Science, vol. 2477. Springer-Verlag, 102--116.

\bibitem[\protect\citeauthoryear{Wegbreit}{Wegbreit}{1975}]{Wegbreit75}
{\sc Wegbreit, B.} 1975.
\newblock {M}echanical {P}rogram {A}nalysis.
\newblock {\em Communications of the ACM\/}~{\em 18,\/}~9.

\end{thebibliography}
%% \end{small}

\newpage

\appendix

\section{The Abstract Interpretation Framework}
\label{sec:absframe}

Abstract interpretation \cite{cousots-lp} is a framework for static
analysis. Execution of the program on a concrete domain is simulated
in an abstract domain, simpler than the former one. Both domains must
be lattices, $\langle \mathcal{P}(\Sigma), \subseteq \rangle$ and
$\langle \Delta, \sqsubseteq \rangle$. To go from one to another we
use a pair of functions, called \emph{abstraction} $\alpha:
\mathcal{P}(\Sigma) \to \Delta$ and \emph{concretization} $\gamma:
\Delta \to \mathcal{P}(\Sigma)$, which should form a Galois
connection: 
$$\langle \mathcal{P}(\Sigma), \subseteq \rangle \galois{\alpha}{\gamma}
\langle \Delta, \sqsubseteq \rangle \text{ if and only if }
\alpha(x) \sqsubseteq y \iff x \subseteq
\gamma(y)$$
Intuitively $\alpha(\sigma)$ generates the smallest
element in $\Delta$ that contains all the elements in $\sigma$, and
$\gamma(\delta)$ computes all the concrete elements represented by
$\delta$.

The methodology is very general, so we focus specifically on the
% For CoRR
% PLAI~\cite{abs-int-naclp89,ai-jlp} framework.  
PLAI \\ \cite{abs-int-naclp89,ai-jlp} framework.  
The PLAI algorithm
abstracts execution \textsc{and-or} trees similarly
to~\cite{DBLP:journals/jlp/Bruynooghe91} but represents the abstract
executions \emph{implicitly} and computes fixpoints efficiently using
memo tables, dependency tracking, etc.  The procedure is
\emph{generic} (\emph{parametric}) in the sense
that it factors out the abstraction of program execution flow (the
execution and-or trees), which is common to many different analyses,
from other (mainly data-related) abstractions, which are more
application-specific, and which are encoded as one or more
\emph{abstract domains}. It is also goal dependent: it takes as input
a pair $(L,\lambda_c)$ representing a predicate along with an
abstraction of the call patterns (in the chosen \emph{abstract
  domain}) and produces an abstraction $\lambda_o$ which
overapproximates the possible outputs, as well as all different
call/success pattern pairs for all called predicates in all paths in
the program and the corresponding abstract information at all other
program points, for all procedure versions. 
This algorithm is the basis of the PLAI abstract
analyzer found in \ciaopp~\cite{hermenegildo11:ciao-design-tplp},
where we have integrated a working implementation of the proposed
resource analysis. In PLAI, abstract domains are pluggable units which
need to define implementations of $\sqsubseteq$, least upper bound
($\sqcup$), bottom ($\bot$), and a number of other operations related
to predicate calls and successes.
% call and return.

For any clause $h \imp q_1, \dots, q_n.$, let $\lambda_i$ and
$\lambda_{i+1}$ be the abstract substitutions to the left and to the
right of literal $q_i$, and $\lambda_{call\_i}$ and
$\lambda_{success\_i}$ their projections onto the variables of $q_i$
respectively.  $\lambda_1$ and $\lambda_{n+1}$ are the \emph{entry}
and \emph{exit} substitutions of the clause respectively, denoted also
as $\beta_{entry}$ and $\beta_{exit}$. We can show this graphically as
follows:

\begin{center}
\begin{tabular}{cc}
\begin{tikzpicture}[scale=0.7]
\node (call) at (-1.5, 1.5) {$\lambda_{call}$};
\node (p) at (0,1.5) {$p$};
\node (success) at (1.5, 1.5) {$\lambda_{success}$};
\node (b1e) at (-4.5,0) {$\beta_{1,entry}$};
\node (h1) at (-3,0) {$h_1$};
\node (b1x) at (-1.5,0) {$\beta_{1,exit}$};
\node (dots) at (0,0) {$\dots$};
\node (bme) at (1.5,0) {$\beta_{m,entry}$};
\node (hm) at (3,0) {$h_m$};
\node (bme) at (4.5,0) {$\beta_{m,exit}$};
\draw (p) -- (h1);
\draw (p) -- (hm);
\end{tikzpicture}
&
\begin{tikzpicture}[scale=.7]
\node (entry) at (-2, 1) {$\beta_{entry}$};
\node (h) at (0,1) {$h$};
\node (exit) at (2, 1) {$\beta_{exit}$};
\node (l1) at (-3,0) {$\lambda_1$};
\node (p1) at (-2,0) {$p_1$};
\node (l2) at (-1,0) {$\lambda_2$};
\node (dots) at (0,0) {$\dots$};
\node (ln) at (1,0) {$\lambda_n$};
\node (pn) at (2,0) {$p_n$};
\node (ln1) at (3,0) {$\lambda_{n+1}$};
\draw (h) -- (p1);
\draw (h) -- (pn);
\end{tikzpicture}
\end{tabular}
\end{center}

To compute $\lambda_{success}$ from $\lambda_{call}$ of a generic
(sub)goal $p(\bar{x})$ with predicate $p$:
\begin{enumerate}
\item Generate a $\beta_{entry\_i}$ from $\lambda_{call}$ for each of
  the $m$ clauses $C_i$ defining the predicate $p$. This transfers the unification of
  the subgoal and head variables into $\Delta$.
\item For each clause $C_i$, compute $\beta_{exit\_i}$ from
  $\beta_{entry\_i}$, and then project $\beta_{exit\_i}$ back again onto the subgoal
  variables, obtaining $\lambda'_i$.
\item Aggregate all the exit substitutions using the least upper
  bound, 
% $\lambda_{success} = \bigcup\limits_{i=1}^{m} \lambda'_i$.
$\lambda_{success} = \bigsqcup\limits_{i=1}^{m} \lambda'_i$.
\end{enumerate}

Computing $\beta_{exit}$ from $\beta_{entry}$ is straightforward: set
$\beta_{entry}$ as $\lambda_1$. Then, project it onto the variables
appearing in the call to the first literal $q_1$,
obtaining 
% . This way
$\lambda_{call\_1}$ 
% is obtained 
for $q_1$, and compute
$\lambda_{success\_1}$ from it
% from which $\lambda_{success\_1}$ can be computed 
using the procedure mentioned
above. Now $\lambda_1$ is integrated with this success substitution,
referred to as \emph{extending} $\lambda_1$ with
$\lambda_{success\_1}$. The result is set as $\lambda_2$, for which the
same series of steps is performed with respect to the second literal
$q_2$. The process continues until $\lambda_{n+1}$ is obtained, which is
actually $\beta_{exit}$.

In the process, more than one call substitution may appear for the same
predicate. This is called \emph{multivariance} of
predicates. Furthermore, if the predicate is recursive, 
a fixpoint needs to be computed. To do so, the process 
above is iterated starting from the 
bottom element of the lattice, $\bot$. 
% For CoRR
% \cite{ai-jlp,inc-fixp-sas}
\\ \cite{ai-jlp,inc-fixp-sas}
describe performant algorithms for this purpose, which are implemented
in \ciaopp.

\section{The Abstract Elements, Redux}
\label{sec:sem}

Because of space constraints, in the main part of the paper the
concrete and abstract domains have not been described in full.
In this section we aim to give a more precise definition of both
elements within the framework of abstract interpretation.

In the concrete domain, the \emph{resource usage} of a predicate
\texttt{p} with respect to a set of resources $r_i$ is given by a set
of triples $(\overline{t}, s, r_{p,i})$, where $\overline{t}$ is a
tuple of terms. The interpretation of such set is that for a call to
\texttt{p} with arguments bound to $\overline{t}$, the number of
solutions is exactly $s$ and the resource usage of each $r_i$ is
exactly $r_{p,i}$. Note that $s$ and $r_{p,i}$ are actual values, not
equations or recurrences. The resource usage is computed by adding the
head cost at the point of entering a clause and the literal cost at
the point of calling a
literal in the body, using the usual SLD resolution
semantics. This definition follows closely the one
in~\cite{resource-verif-iclp2010}, but extended to support several
resources and cardinality.

Let $dom(e)$ be the set of tuples of terms $\overline{t}$ for which a
concrete element $e$ has information over its resource usage. We
define $e \sqsubseteq_c e'$ if and only if $dom(e) \subseteq dom(e')$
and for each $\overline{t} \in dom(e)$, $(p(\overline{t}), s, r_{U,i})
= (p(\overline{t}), s', r'_{U,i})$. That is, the set of terms of the
smaller element must be a subset of the larger one, and the
cardinality and resource usage must coincide in the common part of
their domains.

This concrete domain is abstracted in three different ways, to get a
compound domain. Two of them have already been discussed in the 
literature: the non-failure and determinacy analyses. Those components
of the abstract domain correspond to abstracting the set of elements
$\overline{t}$ using a regular type abstract domain and then
summarizing for those elements whether $s = 0$ or $s > 0$ (for the
non-failure domain) and whether $s = 1$ or $s \neq 1$ (for the
determinacy one). The $failed?$ component of the abstract elements
follows closely the non-failure analysis, keeping different
information during the analysis, but with the same result.

For the recurrences part, we perform several abstractions. First of
all, we move from strict values for the number of solutions and
resource usage to value bounds. Thus, the elements are sets of triples
$(\overline{t}, (s_L, s_U), (r_{L,i}, r_{U,i}))$. The ordering
is now given by:
$$
\begin{array}{lcl}
e \sqsubseteq_1 e' & \iff & dom(e) \subseteq dom(e') \\
& \text{and} & \text{for each } \overline{t} \in dom(e), (s_L, s_U) \subseteq (s'_L, s'_U)
               \text{ and } (r_{L,i}, r_{U,i}) \subseteq (r'_{L,i}, r'_{U,i})
\end{array}
$$
The abstraction function in this case is very simple, we just need to
send each value to an interval with it as only point:
$$\alpha_1(\{ (\overline{t}, s, r_{p,i}) \}_{\overline{t}}) =
\{ (\overline{t}, (s, s), (r_{p,i}, r_{p,i})) \}_{\overline{t}}$$

The second abstraction involves summarizing the domain of each
$\alpha_1(e)$ using the sized types abstract domain.  As
discussed in~\cite{sized-types-iclp2013}, a set of terms is described
via sized types using sized type schemas along with a domain $d$ which
tells which are the values of the bound variables which are covered by
the abstract element, and a set of recurrences $r$ which defines the
relations that bound variables must satisfy between them.  When adding
resource usage information, apart from the bounds from sized types we
can refer to new variables: $s_L$ and $s_U$ refer to the upper and
lower bound in the number of solutions, and $v_{resources}$ contains
such variables for each resource in the system.

In this case, it is easier to give the concretization function
to move from an abstract element $e$ to one in the intermediate
abstract domain:
$$\gamma_2(\langle d, (s_L, s_U), v_{res}, r) \rangle) = 
\bigcup_{\overline{t} \, \in \, \gamma_{\text{sized types}}(\langle d, r \rangle)}
( \overline{t}, bound_{(s_L, s_U)}(\overline{t}, r), bound_{v_{res}}(\overline{t}, r) )
$$
where $bound_v(\overline{t}, r)$ returns the upper and lower
\emph{numerical} bounds for the variables $v$ as given in the
recurrences $r$ for the tuple of values $\overline{t}$.  In few words,
$\gamma_2$ takes all the possible tuples of values given by the sized
type we refer to, and computes the cardinality and resource usage of
each of them as given by the recurrence equations.

The intermediate domain and this concretization function allows us to
define an ordering $\sqsubseteq$ in the abstract elements. But, as
stated in the main part of the paper, doing so would entail knowing
whether some recurrences define a set that is larger or smaller than
another one.  This is an undecidable problem, and thus we need to
resort to other checks which, while being correct, are not
complete. In our case, we chose to use a syntactic check.

From $\alpha_1$ we can obtain the corresponding concretization
function $\gamma_1$, and from $\gamma_2$ we can do the same to obtain
an $\alpha_2$. By composition we obtain the abstraction $\alpha_r =
\alpha_2 \cdot \alpha_1$ and concretization $\gamma_r = \gamma_1 \cdot
\gamma_2$ functions that define the Galois connection between concrete
resource usage triples and the abstract domain of recurrence
equations.

As stated before, our complete abstract elements:
$$\langle (s_L, s_U), v_{resources}, failed?, d, r, nf, det \rangle$$
are the combination of that given by $\langle \alpha_r, \gamma_r
\rangle$ with those of non-failure (which give the $failed?$ and $nf$
components) and determinism (which gives the $det$ component), which
abstract information about $s$ over all possible values. For an
abstract element $a$ to be smaller than $b$, it must be smaller in all
of the three domains at the same time.

\end{document}